\documentclass[12pt]{article}

\usepackage{graphicx,epsfig}
\usepackage{cite}

\newcommand{\p}{{\bf p}}
\newcommand{\q}{{\bf q}}

\newcommand{\w}{E({\bf p})}
\newcommand{\W}{E({\bf p}+{\bf q})}

\begin{document}

\begin{center}
{\large {\bf Non-Analytic Vertex Renormalization of a Bose Gas at Finite 
Temperature}} 

\vspace{1.5cm}

G. Metikas$^{1}$ and G. Alber$^{2}$

$^{1}$ Abteilung f\"ur Quantenphysik, Universit\"at Ulm,
Albert-Einstein Allee 11, D-89069, Ulm, Germany

$^{2}$ Institut f\"ur Angewandte Physik, Technische Universit\"at Darmstadt,
D-64289, Darmstadt, Germany

\end{center}

\vspace{3cm}

\begin{abstract}

We derive the flow equations for the symmetry unbroken phase
 of a dilute 3-dimensional Bose gas. We point out that the flow equation
 for the interaction contains parts which are non-analytic at the origin
 of the frequency-momentum space. We examine the way this non-analyticity 
 affects the fixed point of the system of the flow equations and shifts
 the value of the critical exponent for the correlation length closer 
 to the experimental result in comparison with previous work where
 the non-analyticity was neglected. Finally, we emphasize the purely 
 thermal nature of this non-analytic behaviour comparing our approach to a 
 previous work where non-analyticity was studied in the context of  
  renormalization at zero temperature.

\end{abstract}

\newpage

\section{Introduction}

Thermal effective actions are in
general non-local in coordinate space because the temperature-dependent 
Green functions
contain parts which are non-analytic at the origin of the
momentum-frequency space\cite{Kapusta,LeBellac}. In a theory with two
interacting scalar fields, one integrates out one of them to find the
 effective action for the other. At finite temperature, provided the coupling
is weak, one usually proceeds by applying perturbation theory,  
and then making an expansion in powers of frequency and momentum
in order to obtain a local effective Lagrangian. It is this latter expansion
which leads to results which are not uniquely defined but depend on
 the path on the frequency-momentum plane through which the origin is 
approached. For example, when the perturbation is truncated at the self-energy
level, the self-energy is non-analytic at the origin. 
The reason is that the expansion is around a singularity
 \cite{WeldonRules}.

This effect was noticed for the first time by Abrahams and Tsuneto in
the 60's, in the context of BCS theory, while they were studying
time-dependent Ginzburg-Landau theory near zero temperature and near the
critical temperature \cite{Tsuneto}. Later it became clear that 
it is the origin of 
Debye screening and of plasma oscillations in QED \cite{Baym,Petitgirard}.
 These two different
physical phenomena correspond to two different ways of approaching the
origin of the momentum-frequency plane. The effects of the
non-analyticity have also been studied in QED$_{3}$ \cite{Zuk} and    
in QCD \cite{Kalashnikov, WeldonCovariant, Klimov, WeldonQuark,
 Braaten, FrenkelHigh}. The non-analyticity
 is also present in the graviton self-energy 
\cite{RebhanAnalytical, RebhanCollective}
 and in higher-order graviton diagrams \cite{FrenkelHard}.
 Even in the much simpler case of interacting scalars the
 non-analyticity of the self-energy persists
 \cite{Hott,Fujimoto,WeldonMishaps,EvansZero}. In the case of 
interacting scalars, an interesting remark is that, 
whenever the internal
propagators in a loop have different masses, the self-energy
is analytic at the origin \cite{Vokos}. The reason is that the singularity
 is not at the origin anymore, allowing thus a uniquely defined
expansion around the origin.    

This paper is based on the simple observation that an essential step
of the renormalization group method (RG) applied to a theory with a
self-interacting field, is to split the field in slow and fast
components, and integrate out the fast field obtaining thus an
effective action for the slow field \cite{WilsonKogut,CreswickWiegel}.
 Therefore, according to the above
discussion about thermal effective actions, when RG techniques are
 applied in the context of thermal field theory,
 we anticipate that effects originating
in the non-analyticity will arise.

 We choose to examine this aspect of RG in the context of a
 3-dimensional homogeneous self-interacting bosonic gas with weak
 repulsive interactions and discuss its possible physical significance
 in this case. However our analysis and conclusions should hold whenever RG is
 used at finite temperature. This choice of system was motivated by the renewed
 interest in the Bose-Einstein condensation due to its recent
 experimental realization. For the interacting gas, the approach which
 is most often used is that of Bogoliubov. However, this is just a
 mean-field type method and, in principle, one can improve upon it by
 using more sophisticated techniques. One possibility near
 the critical region is the renormalization group \cite{
 StoofBijlsmaRen,Alber,AlberMetikas,Andersen}.

In the case of the homogeneous gas, there is an extra, more important
reason for looking for alternatives to the Bogoliubov approach.
In the critical region, the Bogoliubov theory 
simply does not work because there are fluctuations around
the mean-field that cannot be treated perturbatively. This happens because,
as the temperature approaches the critical temperature $T_{c}$, the thermal
cloud density develops an infrared singularity and thus diverges as
the momentum tends to zero \cite{RSFW,FermiBurnett}.

In section 2, we introduce the basics of the BEC formalism above the
 critical region. We then apply Wilsonian renormalization and derive the
 flow equations for the parameters of the Lagrangian. We point out the 
 non-analytic structure of the RG correction to the interaction term (vertex)
 and follow this non-analyticity as it propagates to the flow equation for
 the interaction.

In section 3, we calculate the non-trivial fixed pont of the system of the 
 flow equations and
 find the critical exponent for the correlation length. We examine the way
 the non-analyticity affects the fixed point.
 We note that taking the non-analyticity into account shifts the value 
 of the critical exponent closer to the experimental result. 

In section 4, we compare our work with \cite{Shankar} where the issue of
 non-analyticity in the context of renormalization is also discussed. We point
 out that the conclusions of \cite{Shankar} hold only at $T=0$ whereas the
 non-analytic behaviour which we are investigating in this paper is purely 
 thermal and vanishes at zero temperature, thus being completely independent
 of the non-analyticity discussed in \cite{Shankar}.

In section 5, we present our conclusions.

\section{Non-analyticity in the uncondensed phase}

When the two-body collisions between bosons are taken to be low-momentum
 or s-wave, the path integral representation of the partition
function of the  homogeneous interacting Bose-gas is given by

\begin{equation}
 Z(\mu,\beta,V,g) \equiv {\rm Tr} e^{-\beta(\hat{H} - \mu \hat{N})} = 
\int \delta[\phi,\phi^*] e^{-S[\phi,\phi^*]} 
\label{partition}
\end{equation}
with the action 
\begin{equation}
S[\phi, \phi^*] =  \frac{1}{\hbar}
 \int_0^{\hbar \beta } d\tau \int_{V} d^D {\bf x}  
\left[ \phi^* (\tau, {\bf x})
[\hbar\frac{\partial}{\partial \tau} - \frac{\hbar^2}{2m} \nabla^{2} - \mu]
 \phi(\tau, {\bf x}) + \frac{1}{2} g |\phi(\tau, {\bf x})|^4
 \right]. 
\label{action}
 \end{equation}

In the low-momentum approximation the interparticle
interaction can be described by the zero-momentum component of the 
Fourier transform of the two-body interaction potential.
Thus, within this approximation, a repulsive,
 short-range potential can be characterized by a positive
 interaction strength $g$. 
In three spatial dimensions this interaction strength is
related to the positive scattering length $a$ of the interparticle
interaction by the familiar relation $g = 4\pi\hbar^2 a/m$. 
The chemical potential is denoted by $\mu$. The case $\mu  < 0$ holds
for $T>T_{c}$ and corresponds to the uncondensed phase whereas $\mu
>0$ describes the condensate which is formed when $T<T_{c}$
\cite{FetterWalecka}. In this paper we will deal only with the uncondensed 
 phase. Starting
from ($\ref{action}$) we can derive the renormalization group equations
for $g$ and $\mu$. This set of coupled differential 
equations can then be used for the study of universal as well as
non-universal properties of the gas 
\cite{StoofBijlsmaRen, Alber, AlberMetikas, Andersen}. In the following we
will set $\hbar=1$.
 
In order to implement the first step of the RG procedure (Kadanoff
transformation), we split the field $\phi(x)$  into
a long-wavelength component $\phi_<(x)$  and into a short-wavelength
 component $\delta \phi_>(x)$. The short-wavelength field involves Fourier
 components which are 
contained only in an infinitesimally thin shell in momentum space of 
thickness $\Lambda(1 - dl)\leq | {\bf p}|\leq \Lambda$ near the cutoff
$\Lambda $, whereas the long-wavelength field has all its Fourier components in
 the sphere whose center is at the origin of the momentum space and its radius
 is $\Lambda(1 - dl)$. We impose no cutoff on the frequency and apply
 the Wilsonian technique of consecutive infinitesimal shell integration only
 to the momentum and not to the frequency \cite{Hertz}.

 We denote the volume of the shell by $\delta V_{{\bf p}}$, 
the volume of the sphere by $V_{{\bf p}}$. The coordinate space volume
 is denoted by $V$. For simplicity we will be referring 
to $\phi_<(x)$ as the lower or slow field and to $\delta \phi_>(x)$ 
as the upper or fast field. Whenever more compact notation is required we 
 will be making use of the following:
\[ x =(\tau,{\bf x}),\hspace{1cm}  p=(p_{0}^{n},{\bf p}) 
  \hspace{1cm} {\rm with} \hspace{1cm}  p_{0}^{n} = 2 \pi n /\beta, \]
  \[ \int dx = \int_{0}^{\beta} d\tau \int_{V} d^{3} {\bf x}, \hspace{1cm}  
   \int dp = \frac{1}{\beta} \; 
 \sum_{n=-\infty}^{\infty} \int \frac{d^{3}{\bf p}}{(2 \pi)^3}. \] 
We integrate out the upper field and are left with an effective
action for the lower field 
\begin{equation}
S_{\rm eff}[\phi_{<}, \phi_{<}^{*}] = S[\phi_{<}, \phi_{<}^{*}] +
\frac{1}{2} {\rm Tr}{\rm  Ln}[1 - \hat{G}^{>} \hat{\Sigma}].
\label{kadaction}
\end{equation}
For details on the derivation of this result and the approximations
involved see \cite{CreswickWiegel}. The ${\rm Tr}$ denotes the trace in 
both the functional and the internal space of 
$ \hat{G}^{>} \hat{\Sigma} $ whereas the ${ \rm tr}$ denotes the trace 
 only in the internal space (see below). The
 hat denotes that the corresponding quantity is a Schwinger-Fock operator 
\cite{SchwingerFock},

 \[ \hat{G}^{>}(\hat{p})  = \left( \begin{array}{cc}
                  B(\hat{p}) & 0 \\
                  0    &   B^{*}(\hat{p})
                  \end{array} \right) \] \\
and \[ \hat{\Sigma}(\hat{x}) = \frac{g}{2} \left( \begin{array}{cc}
                   4 \phi^{*}_{<}(\hat{x}) \phi_{<}(\hat{x})  & 2 
\phi_{<}(\hat{x}) \phi_{<}(\hat{x}) \\
 2 \phi_{<}^{*}(\hat{x}) \phi_{<}^{*}(\hat{x}) &  4 \phi^{*}_{<}(\hat{x}) \phi_{<}(\hat{x})
                  \end{array} \right) \] \\
with
\begin{equation}
B(\hat{p})= B(\hat{p}_{0},\hat{{\bf p}})=\frac{1}{i \hat{p}_{0} + 
E(\hat{{\bf p}})}, \hspace{2cm}  E(\hat{{\bf p}}) = 
\frac{\hat{{\bf p}}^2}{2m} - \mu.
\end{equation}
Note that the expression for ${\hat\Sigma}$ contains the coupling $g$. This
enables us to perform a perturbative expansion over $g$ in
 (\ref{kadaction}) in order to calculate it explicitly. We truncate this
expansion to second order in $g$

\[ {\rm Tr} {\rm Ln} [1-\hat{G}^{>} \hat{\Sigma}] \approx {\rm Tr}
[ -\hat{G}^{>} \hat{\Sigma} - \frac{1}{2} (\hat{G}^{>} \hat{\Sigma})^2]. \]
The first trace is:
\begin{eqnarray}
 { \rm Tr }[ \hat{G}^{>} \hat{\Sigma}]&&= \int dx~\int dp~ 
{\rm tr}[G^{>}(p) \Sigma(x)] \nonumber \\
&& = \int dx~ 
  |\phi_{<} (x) |^2 2g
 \int_{\delta V_{{\bf p}}}
 \frac{ d^{3} {\bf p} }{ (2 \pi )^3 }  [1+2N[\w ]]
\end{eqnarray}
where $N[\w ] = [e^{\beta [\w ]} -1]^{-1}$  
is the Bose-Einstein distribution. We note that the
first trace is quadratic in the modulus of the lower field and can
therefore be interpreted as a correction to the chemical potential

\begin{equation}
d\mu = -  g
 \int_{\delta V_{{\bf p}}}
 \frac{ d^{3} {\bf p} }{ (2 \pi )^3 }  [1+2N[\w ]].
\end{equation}
The second trace is:
\begin{eqnarray}
&& {\rm Tr}[\hat{G}^{>} \hat{\Sigma} \hat{G}^{>} \hat{\Sigma}]
=\int dp \int dk \int dx \int dy \;  e^{i(p-k)(y-x)} \; 
{\rm tr} [ G^{>}(p) \Sigma(x) G^{>}(k) \Sigma(y) ] = \nonumber \\
&& \int dp \int dk \int dx \int dy \;  e^{i(p-k)(y-x)}
 \; \frac{g^2}{4} \nonumber \\
&& \left[ 16\; B(p)\; B(k)\; \phi_{<}^{*}(x)\; \phi_{<}(x)\; \phi_{<}^{*}(y)
\phi_{<}(y)\;  + 4\; B(p)\; B^{*}(k)\; \phi_{<}(x)\; \phi_{<}(x)\;
 \phi_{<}^{*}(y)\; \phi_{<}^{*}(y) \; + \right.
\nonumber \\ 
&& \left. 4\; B^{*}(p)\; B(k)\; \phi_{<}^{*}(x)\;  \phi_{<}^{*}(x)\;
 \phi_{<}(y)\; \phi_{<}(y)\; + 16 \;B^{*}(p)\; B^{*}(k)\; \phi_{<}^{*}(x)\;
 \phi_{<}(x)\; \phi_{<}^{*}(y)\; \phi_{<}(y)\; \right].\nonumber \\
\end{eqnarray} 
In order to simplify the above expression we change variables as follows:
\begin{enumerate}
\item In the second and fourth terms in the square brackets, $p \rightarrow
-p$ and $ k \rightarrow -k$.
\item In the second term, $ x \rightarrow y$ and $ y \rightarrow x$.
\end{enumerate}
The second trace now becomes:
\begin{eqnarray}
&&{\rm Tr} [ \hat{G}^{>} \hat{\Sigma} \hat{G}^{>} \hat{\Sigma}]=
 \int dp \int dk \int dx \int dy \;  \frac{g^2}{4}
 \nonumber \\
&&\left\{ \left[ e^{i(p-k)(y-x)} + e^{i(k-p)(y-x)} \right]\;
 B(p)\; B(k)\; 16\; \phi_{<}^{*}(x)\; \phi_{<}(x)\; \phi_{<}^{*}(y)\;
 \phi_{<}(y)\; \right. \nonumber \\
&&\left. + 2 e^{i(p-k)(y-x)}\; B^{*}(p)\; B(k)\; 4\; \phi_{<}^{*}(x)\;
 \phi_{<}^{*}(x)\; \phi_{<}(y)\; \phi_{<}(y) \right\}. 
\end{eqnarray}
Changing variables again, $k \rightarrow q=k-p$, yields:
\begin{eqnarray}
&&{\rm Tr} [ \hat{G}^{>} \hat{\Sigma} \hat{G}^{>} \hat{\Sigma}]=
 \int dq \int dx \int dy \; \frac{g^2}{4} \nonumber \\
&& \left\{ 2\; e^{-iq(y-x)}\; J_{1}(q)\; 4\; \phi_{<}^{*}(x)\; 
\phi_{<}^{*}(x)\; \phi_{<}(y)\; \phi_{<}(y)\; \right. \nonumber \\
&& \left. +  \left[ e^{-iq(y-x)} + e^{iq(y-x)} \right]\; J_{2}(q)\;  
16\; \phi_{<}^{*}(x)\; \phi_{<}(x)\; \phi_{<}^{*}(y)\; \phi_{<}(y)\; \right\}
\label{inq}
\end{eqnarray}
where
\begin{equation}
J_{1}(q_{0}^{m}, \q )=\int dp \; B^{*}(p) \; B(p+q) 
\hspace{1cm}  {\rm and} \hspace{1cm} 
J_{2}(q_{0}^{m}, \q )=\int dp \; B(p) \; B(p+q).
\end{equation}
We note that ${\bf p}$ is the momentum of the upper field 
(integrated over the infinitesimal shell around the cutoff) whereas ${\bf q}$ 
 is the momentum of the lower field, Fig.\ref{r0}.

It is essential in the RG procedure, and in particular in the Kadanoff
transformation, to recast the effective action obtained after
integrating out the upper field in the form of the original action 
(\ref{action}). The first trace is in a form that can be interpreted as
a correction to the chemical potential. This is not the case however
for the second trace; there are quartic products of fields but,
unlike the four-field coupling term in the original action
(\ref{action}), these are non-local in coordinate space, thus not
allowing the effective action to be recast in the form of the original
action. In other words, though we start from an action containing interactions
 which are local in coordinate space, the RG procedure
generates more general, non-local interactions. 

This is a well-known 
feature of RG, namely to generate extra terms that do not appear in
the original action and have a more general form in comparison to what we
started with \cite{WilsonKogut,Fisher,Cardy}. Provided that these extra terms
 are not relevant they can, in most cases, for the purpose of calculating
 universal properties, be discarded. 

In our case we can Taylor
expand  $\phi_{<}(y)$ around $\phi_{<}(x)$. If we truncate this expansion
 at leading order, $\phi_{<}(y) \approx \phi_{<}(x)$, we remain within
  the family of local interactions we started with. This procedure is 
 usually called derivative expansion and is physically relevant 
only when the lower field is slowly varying both in space and in time.
 
Wilsonian renormalization is compatible with the derivative expansion.
 The reason is that in Wilsonian renormalization we are interested 
in constructing an effective action for the slow field. This compatibility
 can also be seen from a more technical point of view; the derivative 
 expansion of the lower field is equivalent to an expansion of $J_{1}$ and
 $J_{2}$ in powers of $q_{0}$ and ${\bf q}$ (this is easily seen from
 (\ref{inq}) doing integration by parts). This means that truncating
  the derivative expansion at higher than the leading order would give 
  momentum and frequency dependent corrections to the interaction. However 
 such terms are known to be irrelevant (see for example \cite{Fisher},
 page 128) and can therefore be omitted. 

The second trace becomes:
\begin{eqnarray}
&& {\rm Tr} [ \hat{G}^{>} \hat{\Sigma} \hat{G}^{>} \hat{\Sigma}]=
 \int dx  \; \frac{g^2}{4} \nonumber \\
&&  \left\{ \lim_{q
\rightarrow 0}[ 2 e^{iqx} J_{1}(q) ]  \ 4 \phi_{<}^{*}(x) \phi_{<}^{*}(x)
 \phi_{<}(x) \phi_{<}(x) + 
\lim_{q \rightarrow 0}[(e^{iqx} + e^{-iqx}) J_{2}(q) ] \ 
16 \phi_{<}^{*}(x) \phi_{<}(x) \phi^{*}_{<}(x) \phi_{<}(x) 
\right\} \nonumber \\
\end{eqnarray}
which is a local expression that can be interpreted as a correction to
the coupling term of the original action
\begin{equation}
dg = - g^2 \left\{ \lim_{(q_{0}^{m},|\q |) \rightarrow (0,0)}[J_{1}(q_{0}^{m},\q)]
                  + 4 \lim_{(q_{0}^{m},|\q |) \rightarrow
                  (0,0)}[J_{2}(q_{0}^{m}, \q)] \right\}.
\label{dg}
\end{equation}
It is at this point that the non-analyticity enters our discussion.
Because we are at finite temperature, the integrals over frequencies
in $J_{1}$ and $J_{2}$, $I_{1}$ and $I_{2}$ respectively, 
become sums which can be easily calculated when
we turn them into integrals on the complex plane through Poisson summation   

\begin{equation}
J_{1}(q_{0}^{m},{\bf q}) = \int_{\delta V_{{\bf p}}} \frac{d^{3}{\bf p}}{(2
  \pi)^3} \; I_{1}(q_{0}^{m}, {\bf q}, {\bf p}) \hspace{1cm} {\rm and}
  \hspace{1cm}  J_{2}(q_{0}^{m}, {\bf q} ) = 
\int_{\delta V_{{\bf p}}} 
\frac{d^{3}{\bf p}}{(2 \pi)^3} \; I_{2}(q_{0}^{m}, {\bf q}, {\bf p})
\end{equation}
where

\begin{equation}
 I_{1}(q_{0}^{m}, {\bf q}, {\bf p})=
 \frac{1}{\beta} \sum_{n=- \infty}^{ \infty } B^{*}(p_{0}^{n},{\bf p})
B(p_{0}^{n}+q_{0}^{m}, {\bf p+q})) = \frac{1+ N[E(\p )] + N[E(\p +
\q)]}{E(\p + \q) + E(\p) -i q_{0}^{m}},
\end{equation}
\begin{equation}
  I_{2}(q_{0}^{m}, {\bf q}, {\bf p})=
 \frac{1}{\beta} \sum_{n=- \infty}^{ \infty } B(p_{0}^{n}, \p )
B(p_{0}^{n}+q_{0}^{m}, \p + \q )) = \frac{N[E(\p )] - N[E(\p +
\q)]}{E(\p + \q) - E(\p) + i q_{0}^{m}}.
\end{equation}
 We have set $\exp{i \beta q_{0}^{m}} =1$ because $q_{0}^{m}=2 \pi m/
\beta$. In the following, we will suppress the superscript of $q_{0}^{m}$ for
 simplicity.

The first sum, $I_{1}$, is non-vanishing at $T=0$ and is known as the regular
term. The second sum, $I_2$, is purely thermal and is usually called Landau
term in the context of thermal field theory \cite{LeBellac}. 
We observe that the successive limits of $J_{1}(q_{0},\q
)$ coincide, i.e. 

\begin{equation} 
\lim_{q_{0} \rightarrow 0} \lim_{|\q | \rightarrow 0}  J_{1}(q_{0},\q ) =
\int_{\delta V_{{\bf p}}} \frac{d^{3} \p}{(2 \pi)^3} \frac{1}{2 \w }\left[1+2N[\w ] \right]
= \lim_{|\q | \rightarrow 0} \lim_{ q_{0} \rightarrow 0}  J_{1}(q_{0},\q )
\label{J1 limits}
\end{equation}
whereas the successive limits of $J_{2}(q_{0},\q )$ do not

\begin{equation} 
\lim_{q_{0} \rightarrow 0} \lim_{|\q | \rightarrow 0}  J_{2}(q_{0},\q ) = 0
 \neq \lim_{|\q | \rightarrow 0} \lim_{q_{0} \rightarrow 0}  J_{2}(q_{0},\q
) = \int_{\delta V_{{\bf p}}} \frac{d^{3} \p}{(2 \pi)^3} \beta e^{\beta \w }
 N^{2}[\w ] .
\label{J2 limits}
\end{equation}

The reason these two limits do not commute is that $J_{2}$ has a
singularity at the origin of the momentum-frequency space \cite{WeldonRules}. 
Of course, in the evaluation of the above limits, we interchanged the limits 
with both the integration over the momentum $|\p | $ and with the angular 
integration over $\theta$, so our conclusion is not entirely reliable so far. 
In principle, one should perform the integrations over 
$|\p |$ and $\theta$ first, and then take the limit. Unfortunately, 
in our case, the integration over 
$|\p |$ cannot be done analytically. However, we can perform the 
angular integration over $\theta $ analytically before
 evaluating the limits, provided that we split the integral as follows

\begin{equation}
J_{2}(q_{0}, \q ) = \int_{\delta V_{{\bf p}}} \frac{d^{3} \p }{(2 \pi)^3} 
\frac{N[E(\p )]}{E(\p + \q) - E(\p) + i q_{0}}
- \int_{\delta V_{{\bf p}}} \frac{d^{3} \p }{(2 \pi)^3}  
\frac{N[E(\p + \q)]}{E(\p + \q) - E(\p) + i q_{0}}
\label{split} 
\end{equation}
 and perform the change of variables $\p \rightarrow -\p -\q $ in the second
term, eliminating thus the dependence of the Bose-Einstein distribution
on the angle $\theta$ and making the angular integration possible. This 
procedure yields the result

\begin{equation}
J_{2}(q_{0}, \q ) = \int_{\delta V_{{\bf p}}}
 \frac{d^{3} \p }{(2 \pi)^3} \left[ 
\frac{1}{\W - \w + i q_{0}}
+ \frac{1}{\W - \w - i q_{0}} \right] N[\w ].
\end{equation}
It is crucial to note that this change of variables is not permissible
in case we interchange the limit $ \lim_{|\q | \rightarrow 0}
 \lim_{q_{0} \rightarrow 0} $ with the integrations, because this
causes both the terms in (\ref{split}) to diverge. These two divergencies
canceled each other before the change of variables $\p \rightarrow -\p -\q$ 
 \cite{Zuk}. Keeping this remark in mind we now perform the angular
 integration and find

\begin{equation}
J_{2}(q_{0}, \q )= \frac{m}{4 \pi^2} \int_{\Lambda - d\Lambda}^{\Lambda}
 d|\p | \frac{|\p | N[\w ]}{|\q |} \ln{\left[ 
 \frac{(E_{+} - \w )^2 + q_{0}^2}{(E_{-} - \w )^2 + q_{0}^2} \right] }
\label{J2 angular}
\end{equation}
where $E_{+}=E(|{\bf p}|+|{\bf q}|)$ and $E_{-}=E(|{\bf p}|- |{\bf q}|)$.
Instead of just taking the two successive limits in $J_{2}$ as we did before, 
which in the momentum-frequency plane corresponds to approaching the
origin in the direction of the one or the other axis, 
we could approach the origin through
any other curve, for example in the direction of any straight line 
$q_{0}= a |\q |$. Here, of course,
we should not forget that the frequency is discrete whereas the
momentum is continuous. However, for the purpose of better
illuminating the structure of $J_{2}$ around the origin, we shall make
the approximation that the frequency is continuous so that
 $q_{0}= a |\q |$ can hold for any real $a$. Applying this
parameterization to (\ref{J2 angular}) and then taking the
 limit $|\q | \rightarrow 0$ yields

\begin{equation}
\lim_{|\q | \rightarrow 0}  J_{2}(a |\q |, |\q |) = \frac{m}{4 \pi^2}
\int_{\Lambda - d\Lambda}^{\Lambda} 
 d|\p | N[\w ] \frac{2 |\p |^2}{|\p |^2 + m^2 a^2}   
\end{equation}
which reproduces the first limit of (\ref{J2 limits})
 for $a \rightarrow \infty$.
This result was derived from (\ref{J2 angular}) and therefore is also not valid
when the limit $a \rightarrow 0$ is interchanged with the integration
over $|\p |$. However, if we do an integration by parts, we find
\begin{eqnarray}
&&  \lim_{|\q | \rightarrow 0}  J_{2}(a |\q |, |\q |) = 
 \frac{m}{2 \pi^2} 
\left[N[\w ] \frac{|\p |^3}{|\p |^2 + m^2 a^2} \right]_{\Lambda - d
\Lambda}^{\Lambda} \nonumber \\
&& + \frac{1}{2 \pi^2} \int_{\Lambda - d\Lambda}^{\Lambda}
 d|\p | |\p |^2
\left[\frac{|\p |^2}{|\p |^2 + m^2 a^2} \beta e^{\beta \w } N^{2}[\w ]
- \frac{2 m^3 a^2}{[|\p |^2 + m^2 a^2]^2}  N[\w ]  \right]. 
\label{J2 by parts}
\end{eqnarray}
We note that the surface term vanishes, as it is evaluated at the cutoff.
 This result not only reproduces the first limit of (\ref{J2 limits}) for 
$a \rightarrow \infty$
but also agrees with the second limit of (\ref{J2 limits}) for $a \rightarrow
 0$. If we perform the angular integration and apply the same parameterization
 to $J_{1}$, at the limit $|\q | \to 0$, $J_{1}$ is independent of $a$ and 
given by (\ref{J1 limits}). Expressions (\ref{J1 limits}) 
and (\ref{J2 by parts}) are to be substituted in
the correction for the coupling constant (\ref{dg}).

Before we proceed to the second step of the RG formalism, we 
 parameterize the momentum according to $|\p(l)| = \Lambda e^{-l}$. The
 purpose this change of variables serves is simply to make the flow equations
 more elegant.

So far, the flow equation for the chemical potential is:
\begin{equation}
\frac{d\mu}{dl}= -g \frac{\Lambda^3 e^{-3l}}{2 \pi^2} [1+2N[\epsilon_{\Lambda}
e^{-2l} - \mu ]]
\label{dmudl} 
\end{equation}
and the flow equation for the coupling is:
\begin{eqnarray}
&&\hspace{-1cm} 
\frac{dg}{dl} = - g^2 \left\{ \frac{\Lambda^{3} e^{-3l}}{2 \pi^2} 
\frac{1}{2  [\epsilon_{\Lambda} e^{-2l} - \mu]} [1+2 N[\epsilon_{\Lambda}
e^{-2l} - \mu]] + \right. \nonumber \\
&&\hspace{-1cm}
 \left. 4 \frac{\Lambda^{3} e^{-3l}}{2 \pi^2} \left[ \frac{\Lambda^2
      e^{-2l}}{\Lambda^{2} e^{-2l} + m^2 a^2} \beta N[\epsilon_{\Lambda}
e^{-2l} - \mu] [ 1+  N[ \epsilon_{\Lambda} e^{-2l} - \mu]] - \frac{2 m^3
  a^2}{[\Lambda^2 e^{-2l} + m^2 a^2]^2} N[\epsilon_{\Lambda} e^{-2l} - \mu]
   \right] \right\} \nonumber \\
\label{dgdl}
\end{eqnarray}
where $\epsilon_{\Lambda}=\Lambda^2/2m$.

At this point we apply the second step of the RG procedure, namely the trivial
rescaling, whose purpose is to bring the effective action in the form of
the original one. There are two stages, first, we rescale the momentum 
according to $|\q | \rightarrow |\q(l) | = |\q | e^{l}$ in order to
re-establish the original cutoff $\Lambda $. Then we require that the
effective Lagrangian has the same form as the original Lagrangian.
 This induces the trivial rescaling of the parameters of the effective 
Lagrangian.  

\begin{eqnarray}
 V & \to & V(l)= V~e^{-3l}, \nonumber \\
 \beta & \to & \beta(l)=\beta~e^{-2l}, \nonumber \\
 \phi & \to & \phi(l) = \phi~e^{3l/2}, \nonumber \\
 \mu &\to & \mu(l)=(\mu+ \Delta \mu)~e^{2l}, \nonumber \\
 g &\to & g(l)= (g+\Delta g)~e^{-l}.
\label{trivialscaling} 
\end{eqnarray}
The trivial rescaling of $\beta $ implies that the frequency is rescaled
 as $ q_{0} \rightarrow q_{0}(l) e^{-2l} $ and therefore
\begin{equation} 
 a = q_{0} / |\q | \rightarrow a(l) \; e^{-l}.
\label{a trivialscaling}
\end{equation}
Recasting (\ref{dmudl}) and (\ref{dgdl}) in terms of rescaled variables 
 yields the flow equations for the corresponding running quantities
\begin{equation}
\frac{ d\mu (l)}{dl}= 2 \mu(l) - g(l) \frac{\Lambda^3}{2 \pi^2} [ 1+2 N_{l}] 
\label{redmudl}
\end{equation}
and
\begin{equation}
\frac{dg(l)}{dl}=- g(l)- g^2(l) \frac{\Lambda^3}{2 \pi^2}
\left\{\frac{1+2N_{l}}{2[\epsilon_{\Lambda} - \mu(l)]} + 
4 \left[ \frac{\Lambda^2}{\Lambda^2+m^2 a^2(l)} \beta(l) N_{l} [1+N_{l}] -
\frac{2m^3 a^2(l)}{[\Lambda^2+m^2 a^2(l)]^2} N_{l}  \right] \right\}
\label{redgdl}
\end{equation}
where $ N_{l} = [e^{\beta (l) [\epsilon_{\Lambda} - \mu (l)]}-1]^{-1}$ 
is the Bose-Einstein distribution in terms of the rescaled variables.

\section{Fixed Point}

It is important to investigate whether the path-dependence of the flow 
equation for the coupling has any consequences on quantities of physical
interest.

We look at a universal property, the critical exponent for the
correlation length. This is calculated from the coupled
system of (\ref{redmudl}) and (\ref{redgdl}). In fact, in order to have 
an autonomous system, we should also take into account the flow equation
 for the inverse temperature, 
\begin{equation}
\frac{d\beta(l)}{dl}= - 2 \beta (l)
\end{equation}
which is just a differential expression of the trivial scaling of $\beta$ 
(\ref{trivialscaling}).
We observe that, although $\beta $ appears in the equations for $\mu$ and $g$,
these do not couple back to the equation for $\beta $. We also note that the
fixed point for $\beta $ is zero, $\beta_{*}=0$.
 The fact that $\beta_{*} =0$  complicates things because $\beta(l) $ 
appears in the flow equations not only explicitly but also 
 through $N_{l}$. This means that, when we evaluate the fixed point for the 
 system of the flow equations, the 
 right-hand side of (\ref{redmudl}) and (\ref{redgdl}) will diverge, because 
$N_{l}$ diverges for $\beta_{*}=0$. This problem is circumvented when we define
 a scaled running chemical potential $M(l)$ and a scaled running coupling
 constant $\tilde{G}(l)$  such that the set of equations for these new 
 parameters decouples from the equation for $\beta $;

\begin{equation}
M(l) = \beta_{\Lambda} \mu (l) \hspace{1cm}  
 {\rm and} \hspace{1cm} \tilde{G} (l) = \Lambda^{3} \beta_{\Lambda} g(l) /
 b(l)
\end{equation} 
where $\beta_{\Lambda}= m/\Lambda^2$, $\epsilon_{>}=1/2$
 and $b(l)=\beta(l)/\beta_{\Lambda}$ is
the scaled inverse temperature. In terms of these new, dimensionless
 parameters

\begin{eqnarray}
&&\hspace{-1.3cm}
\frac{dM(l)}{dl}=2 M(l) - \frac{1}{2 \pi^2} \tilde{G}(l) b(l) [1+2N_{l}], 
 \nonumber \\
&&\hspace{-1.3cm}
 \frac{d\tilde{G}(l)}{dl}= \tilde{G}(l) -  \frac{1}{2\pi^2} \tilde{G}^{2}(l) 
 b(l) \left\{ \frac{1+ 2N_{l}}{2[\epsilon_{>} - M(l)]} + 4
   \left[\frac{\Lambda^2}{\Lambda^2+ m^2 a^2(l)} b(l) N_{l}[1+N_{l}] 
- \frac{1}{\beta_{\Lambda}} \frac{2 m^3 a^2(l)}{[\Lambda^2+m^2 a^2(l)]^2}
N_{l}  \right] \right\}, \nonumber \\
&&\hspace{-1.3cm}
 \frac{db(l)}{dl}= - 2 b(l).  
\label{scaled system}
\end{eqnarray}
In the neighbourhood of the fixed point, 
the rescaled temperature is high and the approximation 
$N_{l} \approx [\beta(l)[\epsilon_{\Lambda} - \mu(l)]]^{-1} =
 [b(l) [\epsilon_{>} - M(l)]]^{-1} $ holds
\cite{StoofBijlsmaRen}. This yields the equations

\begin{eqnarray}
&&\frac{dM(l)}{dl}=2 M(l) - \frac{1}{2 \pi^2} \tilde{G}(l) \; b(l) \; 
[1 + 2 \frac{1}{b(l)[\epsilon_{>} - M(l)]}], \nonumber \\
&& \frac{d\tilde{G}(l)}{dl}= \tilde{G}(l) -  \frac{1}{2\pi^2} \tilde{G}^{2}(l) 
  \left\{ \frac{b(l)}{2[\epsilon_{>} - M(l)]} + 
          \frac{1}{[\epsilon_{>} - M(l)]^2} + 
  4  \frac{\Lambda^2}{\Lambda^2 + m^2 a^2(l)} \; 
\frac{b(l)}{\epsilon_{>} - M(l)} 
 \right.
\nonumber \\
&&\hspace{1.2cm}  \left.  + 4 \frac{\Lambda^2}{\Lambda^2 + m^2 a^2(l)} \; 
\frac{1}{[\epsilon_{>} - M(l)]^2} - 4 
\frac{2 \Lambda^2 m^2 a^2(l)}{[\Lambda^2 + m^2 a^2(l)]^2} \;
 \frac{1}{\epsilon_{>} - M(l)}  \right\}, \nonumber \\
&& \frac{db(l)}{dl}= - 2 b(l). 
\label{approx system} 
\end{eqnarray}
At the fixed point  $(M_{*},
\tilde{G}_{*}, b_{*}=0)$ the left hand side of (\ref{approx system})
 is zero by definition. The form of the right hand side depends subtly on
 whether $a$ is zero or not as we will see. Near the fixed point, the second 
 term is dominant in the square brackets of the flow equation for the chemical 
 potential in (\ref{approx system}), 
\begin{equation}
\frac{dM(l)}{dl}=2 M(l) - \frac{1}{\pi^2} \tilde{G}(l) 
\frac{1}{\epsilon_{>} - M(l)}.
\end{equation}
 
 For $a=0$, we recall (\ref{a trivialscaling}), 
which means that $a(l)=0$ for any $l$.
 Consequently, the equation for the coupling reduces to
\begin{equation}
 \frac{d\tilde{G}(l)}{dl}= \tilde{G}(l) -  \frac{1}{2\pi^2} \tilde{G}^{2}(l)
\left\{ \frac{b(l)}{2[\epsilon_{>} - M(l)]} + 
          \frac{1}{[\epsilon_{>} - M(l)]^2} + 
 4   \frac{b(l)}{\epsilon_{>} - M(l)} + 4 \frac{1}{[\epsilon_{>}
    - M(l)]^2} \right\}.
\end{equation}
Recalling the trivial scaling of $b$ (\ref{trivialscaling}), 
we see that the second and fourth terms in the curly brackets above
 are dominant near the fixed point,  
\begin{equation}
\frac{d\tilde{G}(l)}{dl}= \tilde{G}(l) -  \frac{1}{2\pi^2} \tilde{G}^{2}(l)
 \frac{5}{[\epsilon_{>} - M(l)]^2}
\end{equation}
and calculate the non-trivial fixed point 
\[ (M_{*},\tilde{G}_{*},b_{*})
=(\frac{1}{12},\frac{5}{4} \frac{\pi^2}{18}, 0).\]
We linearize the system of (\ref{approx system}) around the fixed point 
and find the largest
eigenvalue $\lambda_{+}=1.878$. Therefore the critical exponent for the
correlation length is $\nu=0.532$, 
which agrees with the one found in \cite{StoofBijlsmaRen}.

For $a \neq 0$, according to the trivial scaling of $a$ 
(\ref{a trivialscaling}),
 $ a_{*} = \infty $ and therefore the first and the third terms 
in the curly brackets of (\ref{approx system})
 vanish near the fixed point, as in the case of $a=0$. From the remaining
 terms the fifth vanishes and, significantly, the fourth is also vanishing
 near the fixed point, leaving as dominant contribution only the 
second term. Consequently the flow equation for the coupling reduces to
\begin{equation}
\frac{d\tilde{G}(l)}{dl}= \tilde{G}(l) -  \frac{1}{2\pi^2} \tilde{G}^{2}(l)
 \frac{1}{[\epsilon_{>} - M(l)]^2}
\end{equation}
and the non-trivial fixed point is 
\[ (M_{*},\tilde{G}_{*},b_{*})=(\frac{1}{4},\frac{\pi^2}{8}, 0).
\]
Linearizing around the fixed point we find that $\lambda_{+}=1.561$ and
therefore $\nu = 0.640$. 
It is interesting to note that we could have found the same fixed point and
critical exponent directly from (\ref{approx system}), had we set 
$a = \infty$ and therefore $a(l) = \infty $ for any $l$.

 This situation is similar to what happens, for
 example, in the case of thermal QED for the photon propagator. Because the
 photon self-energy is non-analytic at the origin, different ways of
 approaching the origin lead to different dispersion relations and
 give rise to different types of excitations \cite{Kapusta,LeBellac}.
 For short wavelengths, the dispersion relation is 
 $q_{0}^2 = \q^2 + m_{P}^2$, where 
$ m_{P}^2 = (e^2/2) (T^2/3 + \mu^2/\pi^2)$ is the thermal mass for
 the transverse photons whereas the longitudinal photons do not propagate.
 However, for long wavelengths, the transverse photons have the dispersion 
 $q_{0}^2 = \omega_{P}^2 + \frac{6}{5} \q^2 $ and the longitudinal photons
  have the dispersion $q_{0}^2 = \omega_{P}^2 +\frac{3}{5} \q^2$, where 
 $\omega_{P}^2 = \frac{2}{3} m_{P}^2$ is the plasma frequency at order $e^2$.
 The phenomenon which we are describing here is of the same mathematical
 nature, the difference being that it is occurring not
 in the propagator but in the vertex between four bosons. To be more precise
 it is the vertex graph corresponding to $J_{2}$, Fig.\ref{r0}, that exhibits the same
 singular behaviour as the photon self-energy in QED.
 
Comparing the two critical exponents derived above with the known experimental
value $\nu=0.670$ \cite{Zinn}, it becomes clear that the second procedure gives an 
 improved estimate. In fact, our estimate is better even than 
the value $\nu =0.613$ calculated in \cite{StoofBijlsmaRen} with 
the inclusion of the marginal three-body scattering term in the 
action.

\section{Flow equations and non-analyticity at zero and finite temperature}

Non-analyticity in the context of RG has been discussed before by Shankar 
 in \cite{Shankar}. In this work, the author gives a detailed overview of 
 the RG approach to interacting, non-relativistic fermions in one, two and
 three dimensions and in certain instances (pages 161, 166, 170, 178) 
 refers to the non-analyticity (or lack thereof) 
 which appears in the one-loop RG corrections to the quartic interaction among
 fermions. This is highly reminiscent of the case we are studying, 
the essential difference being that we are dealing with bosons 
instead of fermions. This 
 would render the main point of this paper - the study of a
 non-analyticity in the flow equation for the interaction - rather trivial and
 expected by extending \cite{Shankar} to bosons.

This is not the case however, the non-analyticity we are studying is of
completely different nature from the one studied in \cite{Shankar}. The
 RG calculations in \cite{Shankar}
 are at $T=0$ whereas ours at $T \neq 0$ and the
 non-analyticity we are referring to is essentially thermal and 
 vanishes at $T=0$. To further clarify this point, let us consider the
 ''zero sound'' (ZS) graph which is studied in Eq.(315),
page 161 of \cite{Shankar}.  

\subsection{Zero Temperature Non-Analyticity}

\subsubsection{Zero Sound Integral}

In the zero-sound calculation of \cite{Shankar},
 the following integral appears:

\begin{equation}
S_{1}[\Omega, q] =  \int_{-\Lambda}^{\Lambda} \frac{dk}{2 \pi}
\int_{- \infty}^{\infty}
 \frac{d\omega }{2 \pi} \ 
 \frac{1}{[ i \omega - k ][i (\omega + \Omega) - k - q  ]}
\end{equation}
where $\Omega$, $q$ are the external frequency and momentum and $\omega$, $k$ 
 are the internal frequency and momentum respectively. The external momentum
 is constrained by the same cutoff as the internal momentum, $-\Lambda \leq
 q \leq \Lambda.$

We focus on the integral over $\omega$. When the external momentum and 
 frequency are zero, the integrand has a double pole
 and therefore $S_{1}[0,0]= 0$, Fig.\ref{r1}.

For non-zero external frequency and momentum, the integrand has two single 
poles. Let us assume that $k > 0$. If $k+q > 0$, both poles
are in the lower half-plane and closing the contour from above yields 
$S_{1}[\Omega,q]= 0$. 
However, if $k+q <0$, the two single poles are in different half-planes and
closing the contour either from above or below yields
\[  S_{1}[\Omega, q] = \int_{-\Lambda}^{\Lambda} \frac{dk}{2 \pi} \frac{i}{\Omega + iq}. \] 
Assuming $k<0$ we can argue the same way, so, only when $k$ and $k+q$ have different signs, the $\omega$-integral is non-zero, Fig.\ref{r2}.

Because of the $k$-integration, there is always 
 a range of $k$ values for which $k$ and
 $k+q$ have different signs. Therefore any non-zero $q$ 
results in a non-vanishing  

\[ S_{1}[\Omega,q]= \int_{- \Lambda}^{\Lambda} \frac{dk}{2 \pi} 
\frac{i}{\Omega + i q} [\theta(k) - \theta(k+q)] \]
which depends sensitively on how the limit $ \{ \Omega,q \} \to \{0,0 \}$
 is taken \cite{Shankar},  
\[ \lim_{\Omega \to 0} \lim_{q \to 0} S_{1}[\Omega,q] = 0 \neq  
\lim_{q \to 0} \lim_{\Omega \to 0} S_{1}[\Omega,q]= - 1. \]

\subsubsection{Renormalization Integral}
  
In the RG calculation of \cite{Shankar}, a one-loop correction to the
 interaction is of the form:
 
\begin{equation}
S_{2}[\Omega, q] =   \left[ \int_{-\Lambda}^{-\Lambda + d\Lambda} +
 \int_{\Lambda-d\Lambda}^{\Lambda} \right] \frac{dk}{2 \pi} 
 \int_{- \infty}^{\infty} \frac{d\omega }{2 \pi} \ 
 \frac{1}{[ i \omega - k ][i (\omega + \Omega) - k - q  ]}.
\end{equation}
As before, for zero external frequency and momentum, $S_{2}[0,0]=0$.

 For non-zero external frequency and momentum, we note that
 $S_{2}$ differs from $S_{1}$ only in the range of integration of the 
 momenta, $ k \in [- \Lambda, -\Lambda + d\Lambda ] \cup [ \Lambda - d\Lambda,
 \Lambda]$ and $q \in [ -\Lambda + d\Lambda, \Lambda - d\Lambda]$. This means
 that $|k|>|q|$ and consequently $k,k+q$ have always the same sign 
throughout the integration over $k$. Therefore $S_{2}[\Omega,q]=0$ which 
has no dependence on how the limits of the external frequency and momentum 
are taken \cite{Shankar}. 
  
 The conclusion is that, when performing RG calculations, 
non-analyticities at the origin of external frequency and momentum space
 vanish, even when they are present in the corresponding zero-sound 
calculations.

\subsection{Thermal Non-Analyticity}

\subsubsection{Zero Sound Integral}

Now consider what happens at $T \neq 0$. 
In zero-sound calculations one has to perform the integral
\begin{equation} 
S_{1}^{\rm{T}}[\Omega_{m}, q] =  \int_{-\Lambda}^{\Lambda} \frac{dk}{2 \pi}
 \frac{1}{\beta} \sum_{n=-\infty}^{n=\infty} \ 
 \frac{1}{[ i \omega_{n} - k ][i (\omega_{n} + \Omega_{m}) - k - q  ]}
\end{equation}
where $\omega_{n}=(2 n + 1) \pi / \beta$, $\Omega_{m}=(2m +1) \pi / \beta$ are
the discretized internal and external frequencies respectively. 

When the external frequency and momentum are zero, we find 
\begin{equation}
S_{1}^{\rm{T}}[0,0]= 
- \int_{-\Lambda}^{\Lambda} \frac{dk}{2 \pi} \frac{\beta}{4} 
\cosh^{-2}[\frac{k \beta}{2}] 
\stackrel{\rm{T} \to 0 }{\longrightarrow} -1.
\end{equation}

For non-zero external frequency and momentum, we obtain 
\begin{equation}
S_{1}^{\rm{T}}[\Omega_{m}, q] =  \int_{-\Lambda}^{\Lambda} \frac{dk}{2 \pi}
\frac{i}{\Omega_{m}+iq} \frac{1}{2} \left\{ \tanh{[\frac{k \beta}{2}]}
- \tanh{[\frac{(k+q) \beta}{2}]} \right\}
\end{equation}
which gives the correct zero temperature limit 
$ S_{1}^{\rm{T} \to 0}[\Omega_{m}, q] = S_{1}[\Omega, q]$.
At $T \neq 0$, 
\begin{equation}
\lim_{\Omega_{m} \to 0} \lim_{q \to 0} S_{1}^{\rm{T}}[\Omega_{m}, q]=0
 \stackrel{ \rm{T} \to 0 }{ \longrightarrow } 0,
\end{equation}

\begin{equation}
\lim_{q \to 0} \lim_{\Omega_{m} \to 0} S_{1}^{\rm{T}}[\Omega_{m}, q]=  
- \int_{-\Lambda}^{\Lambda} \frac{dk}{2 \pi} \frac{\beta}{4} 
\cosh^{-2}[\frac{k \beta}{2}] 
\stackrel{\rm{T} \to 0 }{\longrightarrow} -1
\end{equation}
and therefore the non-analyticity of the zero-temperature 
zero-sound calculation persists at finite temperatures, Fig.\ref{r3}.

\subsubsection{Renormalization Integral}

As at $T=0$, the integral appearing in RG calculations, differs from 
 $S_{1}^{\rm{T}}$ only in the range over which the internal and
external momentum are integrated.

 When the external frequency and momentum are zero, we find
\begin{equation}
S_{2}^{\rm{T}}[0,0]= 
 - \left[ \int_{-\Lambda}^{-\Lambda + d\Lambda} +
 \int_{\Lambda-d\Lambda}^{\Lambda} \right] \frac{dk}{2 \pi} \frac{\beta}{4} 
\cosh^{-2}[\frac{k \beta}{2}] 
\stackrel{\rm{T} \to 0 }{\longrightarrow} 0.
\end{equation}

For non-zero external frequency and momentum, we obtain  
\begin{equation} 
S_{2}^{\rm{T}}[\Omega_{m}, q] =  \left[ \int_{-\Lambda}^{-\Lambda + d\Lambda} +
 \int_{\Lambda-d\Lambda}^{\Lambda} \right] \frac{dk}{2 \pi} 
 \frac{1}{\beta} \sum_{n=-\infty}^{\infty} \ 
 \frac{1}{[ i \omega_{n} - k ][i (\omega_{n} + \Omega_{m}) - k - q  ]}.
\end{equation}
Poisson summation yields, Fig.\ref{r3}, 
\begin{equation}
S_{2}^{\rm{T}}[\Omega_{m}, q] =  \left[ \int_{-\Lambda}^{-\Lambda + d\Lambda} +
 \int_{\Lambda-d\Lambda}^{\Lambda} \right] \frac{dk}{2 \pi}  
\frac{i}{\Omega_{m}+iq} \frac{1}{2} \left\{ \tanh{[\frac{k \beta}{2}]}
- \tanh{[\frac{(k+q) \beta}{2}]} \right\}
\end{equation}
which gives the correct zero-temperature limit
\[ S_{2}^{\rm{T} \to 0}[\Omega_{m}, q]=0. \] However, unlike the zero
 temperature renormalization integral, the finite temperature
 renormalization integral is non-analytic, because

\begin{equation}
\lim_{\Omega_{m} \to 0} \lim_{q \to 0} S_{2}^{\rm{T}}[\Omega_{m}, q]=0
 \stackrel{ \rm{T} \to 0 }{ \longrightarrow } 0,
\end{equation}

 \begin{equation}
\lim_{q \to 0} \lim_{\Omega_{m} \to 0} S_{2}^{\rm{T}}[\Omega_{m}, q]=  -
\left[ \int_{-\Lambda}^{-\Lambda + d\Lambda} +
 \int_{\Lambda-d\Lambda}^{\Lambda} \right] \frac{dk}{2 \pi}  \frac{\beta}{4} 
\cosh^{-2}[\frac{k \beta}{2}] 
\stackrel{\rm{T} \to 0 }{\longrightarrow} 0.
\end{equation}
Therefore, at finite temperature, the non-analyticity of
 the zero-sound one-loop graph persists but there is also an extra,
 purely thermal non-analyticity appearing in the finite 
temperature renormalization one-loop graph. 

The zero temperature non-analyticities are due to the splitting of 
 a double pole into two single poles residing in different 
half-planes. The finite temperature non-analyticities are due only
 to the splitting of a double pole and appear even if the resulting
 two single poles reside in the same half-plane. Therefore it is
 only natural that extra non-analyticities appear at finite 
temperature in addition to those existing at zero-temperature. 

\section{Conclusions}

We have investigated thermal non-analyticities at the origin of
 the momentum-frequency space in the context of Wilsonian
 renormalization. We have shown that taking them into account leads 
 to an improved estimate for the critical exponent of the
 correlation length, when approaching the critical region 
 from the symmetry unbroken phase. It is well-known that the
 $\epsilon$-expansion gives values closer to the experiment than 
 ours, however one should remember that the $\epsilon$-expansion is
 asymptotic \cite{Zinn}. It is therefore crucial to find safer ways of
 calculating the critical exponent, such as the improved 
 momentum-shell method which was used here.

 When approaching the critical region from the symmetry-broken phase
 there is an excellent estimate $\nu=0.685$ \cite{StoofBijlsmaRen}. It
 may be possible to improve this value by taking into account the 
 non-analyticity as we did for the approach from the symmetry 
 unbroken phase. Work on this issue is currently under progress.

 Finally, we have pointed out that when one applies Wilsonian renormalization
 at finite temperature, one may encounter non-analytic behaviour 
 which is different in nature from the non-analyticity
 (or lack thereof) discussed in \cite{Shankar}.

\section*{Acknowledgement}

The authors wish to thank W. Wonneberger 
for bringing reference \cite{Shankar} to our attention. 
G.M. wishes to thank I.J.R.  Aitchison for many
 discussions on the subtleties of thermal field theory
 as well as  V. Yudson for a useful discussion on renormalization
 and statistical physics. This work was supported by the Deutsche
 Forschungsgemeinschaft within the Forschergruppe ``Quantengase''.  


\begin{center}

\newpage

\begin{figure}
\begin{center}
\includegraphics[angle=270,width=4in]{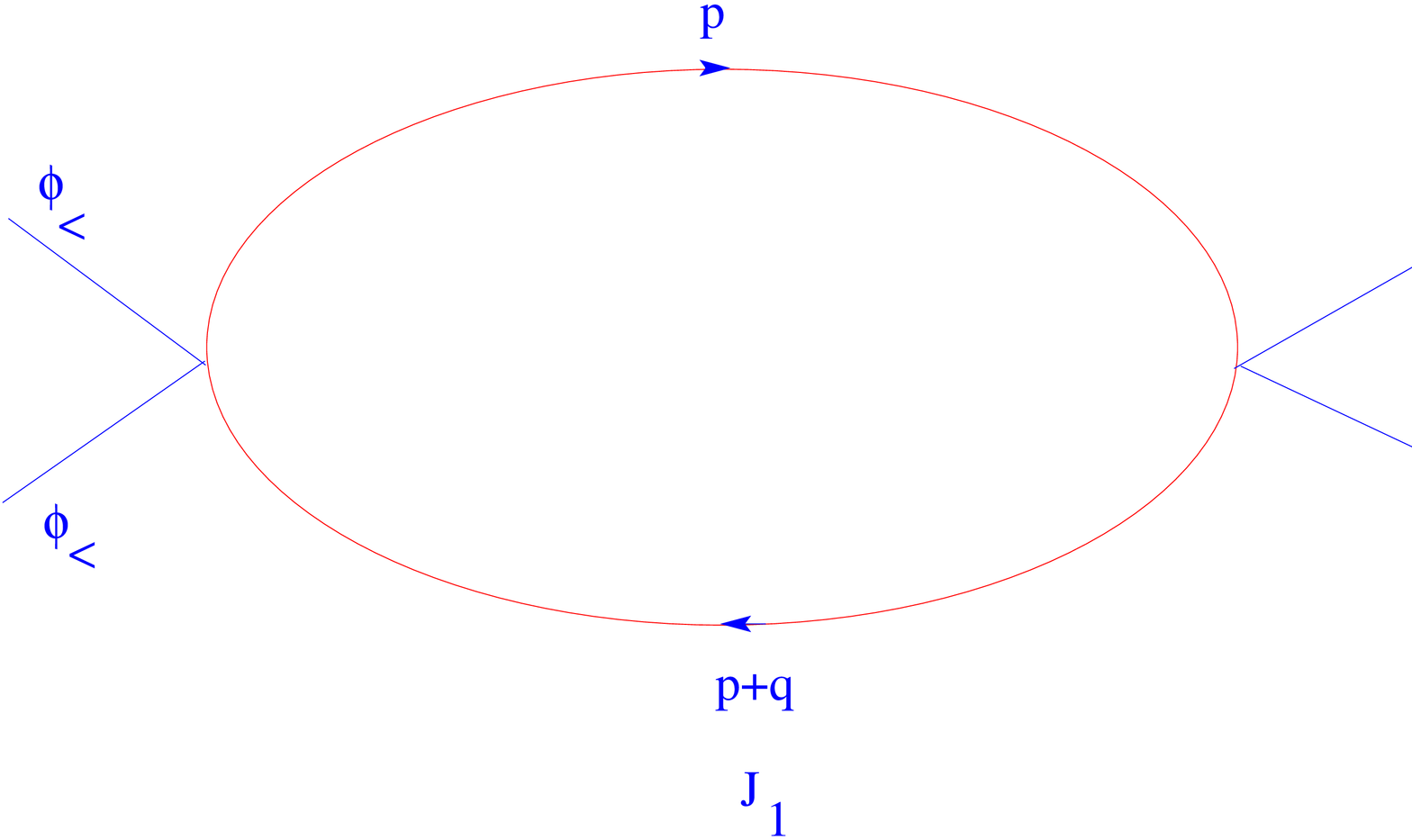}
\includegraphics[angle=270,width=4in]{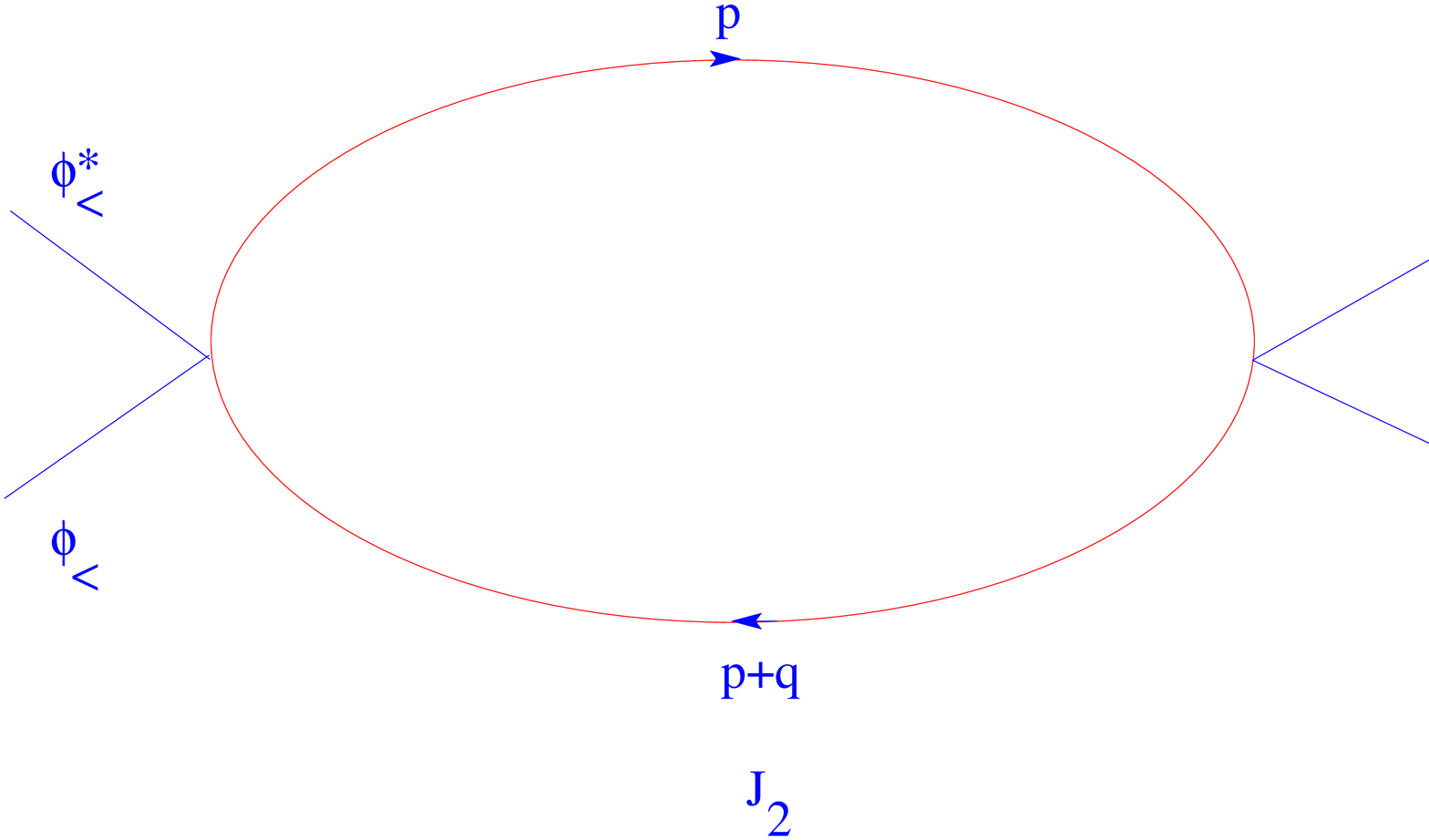}
\caption {The two contributions of the RG correction to the interaction g,
 $J_{1}$ and $J_{2}$; $p$ is the frequency-momentum of the upper field 
 (the momentum ${\bf p}$ is integrated over the infinitesimal shell around the 
 cutoff), $q$ is the frequency-momentum of the lower field. \label{r0}}
\end{center}
\end{figure}

\newpage

\begin{figure}
\begin{center}
\includegraphics[angle=270,width=6in]{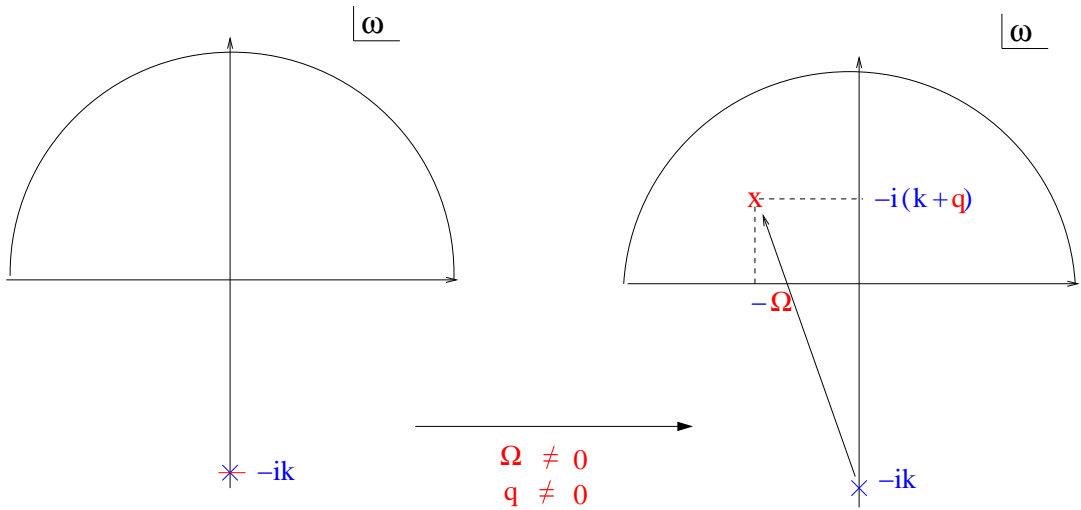}
\caption {
 The integrand of the $\omega$-integration has one double pole and
  the integral is zero. Any non-zero external momentum splits 
 the double pole into two single poles. There is always a range of $k$ 
 for which the two single poles reside in different half-planes and
 consequently the integral is non-vanishing.  \label{r1}}
\end{center}
\end{figure}

\newpage

\begin{figure}
\begin{center}
\includegraphics[angle=270,width=6in]{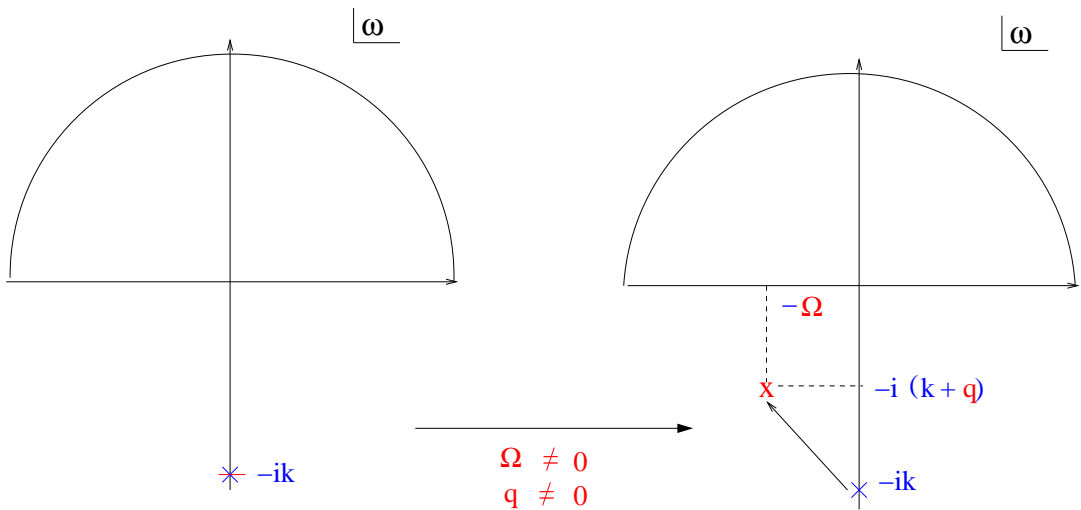}
\caption { The integrand of the $\omega$-integration has one double pole and
  the integral is zero. The introduction of non-zero external momentum splits
 the double pole into two single poles. However, because the $k$-integration 
 is over the infinitesimal shell near the cutoff and the external momentum $q$ 
  takes value below the infinitesimal shell, the two single poles are always
 on the same half-plane and the integral remains zero. \label{r2}}
\end{center}
\end{figure}

\newpage

\begin{figure}
\begin{center}
\vspace{2cm}
\includegraphics[angle=270,width=5in]{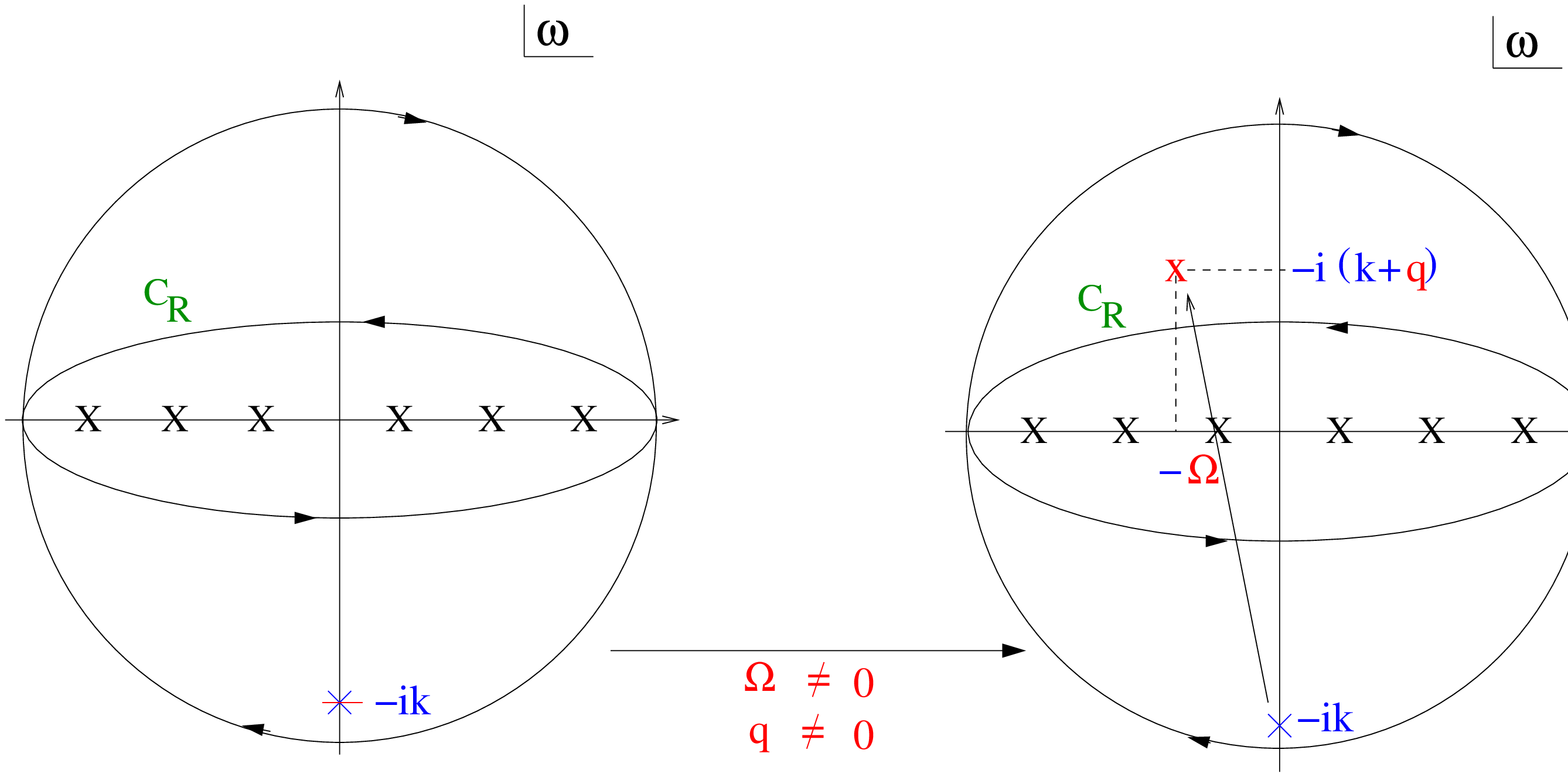}
\includegraphics[angle=270,width=5in]{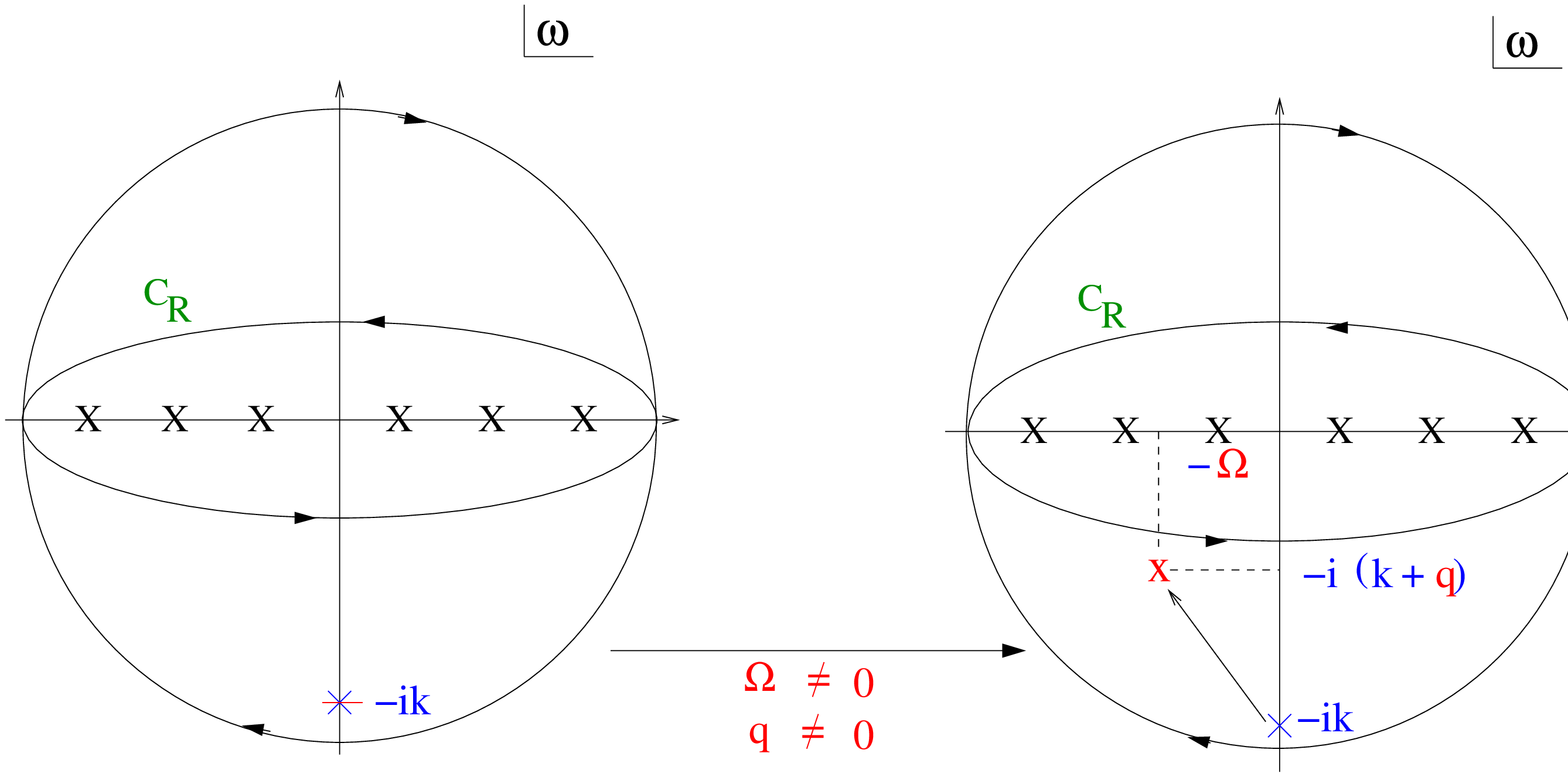}
\vspace{-0.5cm}
\caption { 
  The integrand of the $\omega$-integration has one double pole. The 
introduction of non-zero external momentum splits the double pole 
into two single poles and the integral is non-vanishing regardless
 of whether the two single poles are on different half-planes or not.
 \label{r3}}
\end{center}
\end{figure}
 
\end{center}

\end{document}